\documentclass[a4paper,11pt]{article}
\usepackage{pos}

\usepackage{siunitx}
\usepackage{slashed}
\usepackage{graphicx}
\usepackage{booktabs,multicol,multirow}
\usepackage[shortlabels]{enumitem}
\usepackage{hyperref}
\usepackage[capitalize]{cleveref}
\usepackage{xspace}
\usepackage[dvipsnames]{xcolor}
\usepackage{soul}
\usepackage[textsize=tiny]{todonotes}
\usepackage{subcaption}
\usepackage[normalem]{ulem}
\usepackage{nicefrac}
\usepackage{bm}
\usepackage{makecell}
\usepackage[makeroom]{cancel}

\usepackage{heppennames2}
\usepackage{tikz}
\usepackage[compat=1.1.0]{tikz-feynhand}
\usepackage{xpatch}
\makeatletter
\xpatchcmd\@HepConStyle
 {\edef\@upcode{\updefault}}
 {\ifdefined\shapedefault\edef\@upcode{\shapedefault}\else\edef\@upcode{\updefault}\fi}
 {}{}
\makeatother

\newcommand{\SM}{{\ensuremath{\text{SM}}}\xspace}

\newcommand{\cm}{c.m.\xspace\ }

\newcommand{\lamNP}{\ensuremath{\Lambda_{\text{NP}}}\xspace}

\newcommand{\Ztwo}{\ensuremath{\mathbb{Z}_2}}

\newcommand{\order}[1]{\ensuremath{\mathcal{O}(#1)}\xspace}

\newcommand{\lamhhh}{\ensuremath{\lambda_{\Ph\Ph\Ph}}\xspace}
\newcommand{\klam}{\ensuremath{\kappa_\lambda}\xspace}
\newcommand{\kala}{$\kappa_\lambda$\xspace}
\newcommand{\Zh}{\ensuremath{\PZ\PSh}\xspace}
\newcommand{\sZh}[1]{\ensuremath{\sigma_{\PZ\PSh}^{#1}}\xspace}
\newcommand{\kZh}[1]{\ensuremath{\kappa_{\PZ\PSh}^{#1}}\xspace}
\newcommand{\fcc}[1]{FCC-ee$_{#1}$}
\newcommand{\epem}{\ensuremath{\Pep\Pem}\xspace}

\newcommand{\gev}{\,\, \mathrm{GeV}}

\newcommand{\ie}{\textit{i.e.}\xspace}
\newcommand{\eg}{\textit{e.g.}\xspace}
\newcommand{\cf}{\textit{cf.}\xspace}

\newcommand{\hepfit}{\texttt{HEPfit}\xspace}
\newcommand{\bat}{\texttt{BAT}\xspace}

\newcommand{\NPI}[1]{\ensuremath{\varepsilon_{\text{theo}}^{#1}}\xspace}

\renewcommand{\logo}{\relax}

\title{
Assessing uncertainties in the determination of the trilinear Higgs self-coupling from single-Higgs observables
}
\ShortTitle{Uncertainties in the determination of \klam from single-Higgs observables}

\author[a]{Henning Bahl}
\author[b]{Philip Bechtle}
\author[c]{Johannes Braathen}
\author[d]{Sven Heinemeyer}
\author[c]{Jenny List}
\author*[b]{Murillo Vellasco}
\author[c,e]{Georg Weiglein}

\affiliation[a]{Institute for Theoretical Physics (ITP), Universit\"at Heidelberg,\\
Philosophenweg 16, 69120 Heidelberg, Germany}

\affiliation[b]{Physikalisches Institut, Rheinische Friedrich-Wilhelms-Universit\"at Bonn,\\
Nussallee 12, Bonn, 53115, NRW, Germany}

\affiliation[c]{Deutsches Elektronen-Synchrotron DESY,\\
Notkestr.~85, 22607 Hamburg, Germany}

\affiliation[d]{Instituto de F\'isica Te\'orica (UAM/CSIC), Universidad Aut\'onoma de Madrid,\\
Cantoblanco, 28049, Madrid, Spain}

\affiliation[e]{Institut f\"ur Theoretische Physik, Universit\"at Hamburg,\\
Luruper Chaussee 149, 22761 Hamburg, Germany}

\emailAdd{bahl@thphys.uni-heidelberg.de}
\emailAdd{bechtle@physik.uni-bonn.de}
\emailAdd{johannes.braathen@desy.de}
\emailAdd{sven.heinemeyer@cern.ch}
\emailAdd{jenny.list@desy.de}
\emailAdd{murillo.vellasco@uni-bonn.de}
\emailAdd{georg.weiglein@desy.de}

\abstract{Circular $\epem$ colliders operating at energies below the di-Higgs production threshold can provide information on the trilinear Higgs self-coupling $\lamhhh$ via its loop contributions to single Higgs production processes and electroweak precision observables. We investigate how well a non-SM value of $\lamhhh$ can be determined indirectly via its loop contributions within a global EFT fit. Using an inert doublet extension of the SM Higgs sector as an example for a scenario of physics beyond the SM that could be realised in nature, we find that theoretical uncertainties related to the treatment of loop contributions and the truncation of the EFT expansion, which are usually neglected, play an important role in determining the sensitivity to $\lamhhh$ in a global fit. The results obtained from such an indirect determination of $\lamhhh$ without taking these additional uncertainties into account would be too optimistic, leading to an artificially high resulting precision for $\lamhhh$. They could therefore be misleading in the quest to precisely identify the underlying physics of electroweak symmetry breaking.  
} 

\FullConference{The European Physical Society Conference on High Energy Physics (EPS-HEP2025)\\
7-11 July 2025\\
Marseille, France\\}

\begin{document}

\begin{flushright}
\texttt{DESY-25-146}\\
\texttt{BONN-TH-2025-33}\\
\texttt{IFT–UAM/CSIC-25-152}
\end{flushright}

\maketitle

\setlength\abovedisplayskip{2pt}%
\setlength\belowdisplayskip{2pt}%
\setlength\abovedisplayshortskip{2pt}%
\setlength\belowdisplayshortskip{2pt}%
\setlength{\intextsep}{3pt}

\vspace{-1mm}
%%%%%%%%%%%%%%%%%%%%%%%%%%%%%%%%%%%%%%%%%%%%%%%%%%%%%%
%%%%%%%%%%%%%%%%%%%%%%%%%%%%%%%%%%%%%%%%%%%%%%%%%%%%%%
\section{Introduction}
\label{sec:intro}
\vspace{-0.4cm}

% %%%%%%%%%%%%%%%%%%%%%%%%%%%%%%%%%%%%%%%%%%%%%%%%%%%%%%
% %%%%%%%%%%%%%%%%%%%%%%%%%%%%%%%%%%%%%%%%%%%%%%%%%%%%%%

An \epem Higgs factory has been identified as the highest priority for the next flagship collider at CERN in the 2020 update of the European Strategy for Particle Physics~\cite{EuropeanStrategyGroup:2020pow}. Proposals for such a machine include circular and linear colliders, which mostly differ by their luminosities and achievable centre-of-mass (c.m.)\ energies. While circular accelerators such as the \fcc{} or CEPC can achieve high luminosities at lower \cm energies, linear colliders such as the Linear Collider Facility (LCF) at CERN can reach much higher \cm energies~\cite{LinearColliderVision:2025hlt}. This difference is particularly significant in the extraction of the trilinear Higgs-boson self-coupling \lamhhh, a key parameter whose determination will yield important information on the shape of the Higgs potential, the nature of electroweak symmetry breaking (EWSB) and the electroweak phase transition (EWPT). Crucially, only linear colliders operating at \cm energies of at least 500~GeV will have access to di-Higgs production processes in $\epem$ collisions (at a $\gamma\gamma$ collider di-Higgs production can already be~accessed at a \cm energy of about 280~GeV~\cite{Barklow:2023ess,Berger:2025ijd,LinearColliderVision:2025hlt}), such as $\epem\to \Zh\Ph$, in which \lamhhh enters at leading order (throughout this work, $\Ph$ denotes the Higgs boson observed with $m_{\Ph}\approx125$~GeV). On the other hand, circular colliders only have access to the indirect effects of \lamhhh via its loop-level contributions to single-Higgs observables, as well as via its effects on electroweak precision observables (EWPOs) from the two-loop level onwards. In the present work, we focus on the loop-level extraction of \lamhhh by means of a global EFT fit, and in particular on the associated theoretical uncertainties. Such global EFT fits have been the focus of several studies to date, see \eg Refs.~\cite{deBlas:2019rxi,deBlas:2022ofj,Maura:2025rcv,terHoeve:2025omu}.

In contrast to previous works, we do not assume that future measurements will show no deviations from the SM; instead, we select a certain BSM framework and generate inputs for the global EFT fits as if this scenario was realised in nature. Concretely, we consider the Inert Doublet Model (IDM)~\cite{Deshpande:1977rw,Barbieri:2006dq} as an example of a simple and well-motivated BSM scenario. The IDM is a two-Higgs-doublet extension of the SM that is invariant under an additional unbroken $\Ztwo$ symmetry, under which one of the Higgs doublets is odd (and thus does not acquire a vacuum expectation value, VEV) while the other, along with all other SM fields, is even; see Refs.~\cite{Aiko:2023nqj,Braathen:2024ckk} for our conventions. In this model a strong first-order electroweak phase transition (FOEWPT) can occur, and its $\Ztwo$ symmetry leads to a natural dark matter candidate. Crucially, however, mass splittings within the model lead to large loop contributions involving the additional Higgs bosons of the extended Higgs sector, which result in large higher-order corrections to \lamhhh~\cite{Kanemura:2016sos,Braathen:2019pxr,Braathen:2019zoh,Aiko:2023nqj},
\footnote{These types of effects are not exclusive to the IDM; originally found for Two-Higgs-Doublet Models~\cite{Kanemura:2002vm,Kanemura:2004mg}, they are now known to occur for a wide range of models with extended scalar sectors~\cite{Aoki:2012jj,Kanemura:2015fra,Kanemura:2015mxa,Kanemura:2016lkz,Kanemura:2017wtm,Kanemura:2017gbi,Chiang:2018xpl,Senaha:2018xek,Braathen:2019pxr,Kanemura:2019slf,Braathen:2019zoh,Bahl:2022jnx,Bahl:2022gqg,Bahl:2023eau,Bahl:2025wzj,Braathen:2025qxf} and/or with classical scale invariance~\cite{Hashino:2015nxa,Braathen:2020vwo}. Although, for concreteness, in this work we only consider the IDM, our findings can be generalised to a larger set of models.} while the single-Higgs couplings remain close to their SM counterparts~\cite{Bahl:2026abc}.

\vspace{-5.mm}
%%%%%%%%%%%%%%%%%%%%%%%%%%%%%%%%%%%%%%%%%%%%%%%%%%%%%%
%%%%%%%%%%%%%%%%%%%%%%%%%%%%%%%%%%%%%%%%%%%%%%%%%%%%%%
\section{Global fit methodology}
\label{sec:method}
\vspace{-0.4cm}

In this work, we investigate how well a non-SM value of \klam can be determined by its loop-level contributions in a global EFT fit and in particular the impact of theory uncertainties on this determination. To that end, we make use of the \hepfit C++ package~\cite{DeBlas:2019ehy}, which allows one to perform global fits including direct and indirect constraints for a variety of New Physics scenarios, including the Standard Model Effective Field Theory (SMEFT). In particular, we employ the same setup as used in Ref.~\cite{deBlas:2022ofj} and perform fits using pseudo-data at the \fcc{} (including all proposed \cm configurations up to 365~GeV) and the HL-LHC, including single-Higgs measurements, EWPOs, and di-boson measurements (see Ref.~\cite{deBlas:2022ofj} and references therein for a full description of the assumed collider luminosities and the expected precision for each observable). The fitting procedure is based on Markov Chain Monte Carlo methods implemented in the \bat~package~\cite{CALDWELL20092197,Beaujean_2015}, which is integrated into \hepfit. However, instead of using the SM predictions as the input data for the fits, we use the IDM predictions (for given benchmark scenarios) as the central values for each fit observable, as an example of a New Physics scenario that could be realised in nature. In the following section, we describe in detail the evaluation of the single-Higgs observables, where we compare the prediction in the considered full model, the IDM, with the SMEFT description.

\vspace{-5.5mm}
%%%%%%%%%%%%%%%%%%%%%%%%%%%%%%%%%%%%%%%%%%%%%%%%%%%%%%
%%%%%%%%%%%%%%%%%%%%%%%%%%%%%%%%%%%%%%%%%%%%%%%%%%%%%%
\section{Calculation of the \texorpdfstring{\Zh}{Zh} cross-section and other single-Higgs observables}
\label{sec:sigmaZh}
\vspace{-0.35cm}

\begin{figure}[htpb]
    \centering
    \begin{subfigure}{0.32\textwidth}
        \centering
        \resizebox{!}{2.6cm}{% \resizebox{2.5cm}{!}{%
\begin{tikzpicture}[scale=1.0]
	\setlength{\feynhandlinesize}{1.0pt}
	\setlength{\feynhanddotsize}{1.5mm}
	\begin{feynhand}
		\vertex [particle] (em) at (-2,+1.5) {\large ${e^{-}}$};
		\vertex [particle] (ep) at (-2,-1.5) {\large ${e^{+}}$};
		
		\vertex [particle] (Z) at (+2.5,+1.5) {\large ${Z}$};
		
		\vertex [particle] (h) at (+2.5,-1.5) {\large ${h}$};
		
		\vertex (veeZ) at  (-0.75,+0);
		\vertex (vZZh1) at (+0.5 ,+0);
		\vertex (vZZh2) at (+1.5 ,+0.5);
		\vertex (vhhh) at  (+1.5 ,-0.5);
		
		\propagator [fermion] (em) to (veeZ);
		\propagator [fermion] (veeZ) to (ep);
		
		\propagator [boson] (veeZ) to [edge label={\large ${Z}$}] (vZZh1);
		\propagator [boson] (vZZh1) to (vZZh2);
		
		\propagator [boson] (vZZh2) to (Z);
		
		\propagator [scalar] (vZZh1) to [edge label'=${h}$] (vhhh);
		\propagator [scalar] (vZZh2) to [edge label=${h}$] (vhhh);
		\propagator [scalar] (vhhh) to (h);
		
		\filldraw [red] (vhhh) circle (4pt);
		\vertex [red, below left=0.1mm of vhhh, yshift=-1mm] {\large \kala};
		
	\end{feynhand}
\end{tikzpicture}%
% }
}

        \vspace{-6mm}
        \caption{}
        \label{fig:ZH_klam_diagrams:vertex1}
    \end{subfigure}\hspace{1mm}
    \begin{subfigure}{0.32\textwidth}
        \centering
        \resizebox{!}{2.6cm}{% \resizebox{2.5cm}{!}{%
\begin{tikzpicture}[scale=1.0]
	\setlength{\feynhandlinesize}{1.0pt}
	\setlength{\feynhanddotsize}{1.5mm}
	\begin{feynhand}
		\vertex [particle] (em) at (-2,+1.5) {\large ${e^{-}}$};
		\vertex [particle] (ep) at (-2,-1.5) {\large ${e^{+}}$};
		
		\vertex [particle] (Z) at (+3,+1.5) {\large ${Z}$};
		
		\vertex [particle] (h) at (+3,-1.5) {\large ${h}$};
		
		\vertex (veeZ) at (-0.75,0);
		\vertex (vZZh) at (+1.0,0);
		\vertex (vhhh) at (+2.0,-0.75);
		
		\propagator [fermion] (em) to (veeZ);
		\propagator [fermion] (veeZ) to (ep);
		
		\propagator [boson] (veeZ) to [edge label={\large ${Z}$}] (vZZh);
		\propagator [boson] (vZZh) to (Z);
		\propagator [scalar] (vZZh) to [quarter left, edge label=${h}$] (vhhh);
		\propagator [scalar] (vZZh) to [quarter right, edge label'=${h}$] (vhhh);
		\propagator [scalar] (vhhh) to (h);
		
		\filldraw [red] (vhhh) circle (4pt);
		\vertex [red, below=1.2mm of vhhh] {\large \kala};
		
	\end{feynhand}
\end{tikzpicture}%
% }
}
        
        \vspace{-6mm}
        \caption{}
        \label{fig:ZH_klam_diagrams:vertex2}
    \end{subfigure}\hspace{2mm}
    \begin{subfigure}{0.32\textwidth}
        \centering
        \resizebox{!}{2.6cm}{\resizebox{2.5cm}{!}{%
\begin{tikzpicture}[scale=1.0]
	\setlength{\feynhandlinesize}{1.0pt}
	\setlength{\feynhanddotsize}{1.5mm}
	\begin{feynhand}
		\vertex [particle] (em) at (-2.25,+1.5) {\large ${e^{-}}$};
		\vertex [particle] (ep) at (-2.25,-2) {\large ${e^{+}}$};
		
		\vertex [particle] (Z) at (+3.5,+1.5) {\large ${Z}$};
		
		\vertex [particle] (h) at (+3.5,-2.) {\large ${h}$};
		
		\vertex (veeZ) at (-1.0,0);
		\vertex (vZZh) at (+1.0,0);
		\vertex (vhhh1) at (+1.8,-0.66);
		\vertex (vhhh2) at (+2.7,-1.33);
		
		\propagator [fermion] (em) to (veeZ);
		\propagator [fermion] (ep) to (veeZ);
		
		\propagator [boson] (veeZ) to [edge label={\large ${Z}$}] (vZZh);
		\propagator [boson] (vZZh) to (Z);
		
		\propagator [scalar] (vZZh) to [edge label=\large ${h}$] (vhhh1);
		
		\propagator [scalar] (vhhh1) to [quarter left, looseness=1.5, edge label=\large ${h}$] (vhhh2);
		\propagator [scalar] (vhhh1) to [quarter right, looseness=1.5, edge label'=\large ${h}$] (vhhh2);
		
		\propagator [scalar] (vhhh2) to (h);
		
		\filldraw [red] (vhhh1) circle (4pt);
		\vertex [red, below left=0.2mm of vhhh1] {\large \kala};
		\filldraw [red] (vhhh2) circle (4pt);
		\vertex [red, right=1.3mm of vhhh2] {\large \kala};
		
	\end{feynhand}
\end{tikzpicture}%
}}
        
        \vspace{-6mm}
        \caption{}
        \label{fig:ZH_klam_diagrams:WFR}
    \end{subfigure}
    \vspace{-3mm}
    \caption{One-loop contributions to the process $\epem\to\Zh$ that depend on \klam.}
    \label{fig:ZH_klam_diagrams}
\end{figure}
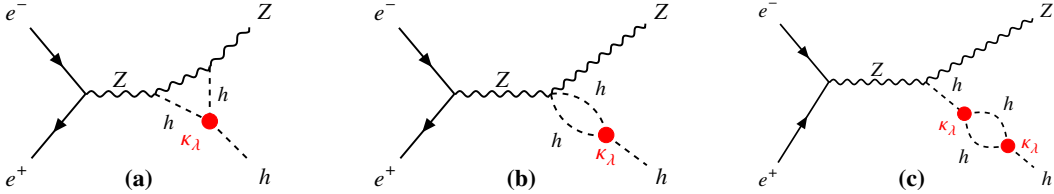

The sensitivity of future circular \epem colliders to \lamhhh, or the corresponding coupling modifier $\klam \equiv {\lamhhh}/{\lamhhh^{\text{SM},(0)}}$ (where $\lamhhh^{\text{SM},(0)}$ is the tree-level prediction in the SM), relies on the measurement of the $\epem\to\Zh$ cross-section. In the SMEFT framework, employing the Warsaw basis \cite{Grzadkowski:2010es}, the leading (dimension-6) BSM contributions to \klam are parametrised in terms of the following operators: $\mathcal{O}_\Phi = (\Phi^\dagger\Phi)^3$, $\mathcal{O}_{\Phi\Box} = (\Phi^\dagger\Phi)\Box(\Phi^\dagger\Phi)$, and
$\mathcal{O}_{\Phi D} = (\Phi^\dagger D^\mu\Phi)^*(\Phi^\dagger D_\mu\Phi)$, where $\mathcal{O}_{\Phi\Box}$ and $\mathcal{O}_{\Phi D}$ modify \klam solely through field redefinitions~\cite{Alasfar:2023xpc}. In terms of the corresponding Wilson coefficients (denoted $C_\Phi$, $C_{\Phi\Box}$, $C_{\Phi D}$) and the New Physics scale $\lamNP$, \klam can be written as~\cite{Alasfar:2023xpc}:\vspace{0.3mm}
\begin{align}
    \text{SMEFT (tree level)}: \qquad
    \klam = 1 - \frac{2v^4}{m_h^2}\frac{C_\Phi }{\lamNP^2}  + \frac{3v^2}{\lamNP^2}\left(C_{\Phi\Box} - \frac{1}{4} C_{\Phi D}\right)\;,
    \label{eq:klam_smeft}
\end{align}
where $\nu$ is the vacuum expectation value of the Higgs field $\Ph$. In this framework, the one-loop diagrams contributing to $\epem\to\Zh$ that include at least one insertion of \klam are shown in \cref{fig:ZH_klam_diagrams}. 

%%%%%%%%%%%%% figure %%%%%%%%%%%%%
%%%%%%%%%%%%% figure %%%%%%%%%%%%%

Following Refs.~\cite{Degrassi:2016wml,Maltoni:2017ims}, the \Zh cross-section can be expressed as a function of \klam as:
\begin{align}
    \sZh{\klam} = Z_{h,\klam} \sigma_\text{LO} (1 + \klam C_1 )\;.
    \label{eq:sigmaZh_smeft}
\end{align}
Here, $\sigma_\text{LO}$ denotes the leading-order cross-section, whereas $\klam\cdot C_1$ corresponds to the interference of the vertex-correction diagrams (\cref{fig:ZH_klam_diagrams:vertex1,fig:ZH_klam_diagrams:vertex2}) with the leading-order contribution. Finally, $Z_{h,\klam}$ denotes the resummed contribution to the wave function renormalisation (WFR) of the Higgs field, corresponding to the diagram in \cref{fig:ZH_klam_diagrams:WFR}, and is given by $Z_{h,\klam}\equiv (1 - \klam^2\delta Z_h)^{-1}$, where $\delta Z_h \simeq - 1.536\times 10^{-3}$~\cite{Degrassi:2016wml}. The ratio of the \Zh cross-section to its SM prediction therefore takes the following form:\vspace{-4mm}
\begin{align}
    \frac{\sZh{\klam}}{\sZh{\klam=1}} = \frac{1-\delta Z_h}{1-\klam^2 \delta Z_h} \frac{1 + \klam C_1}{1 + C_1}\;.
    \label{eq:Zh_full}
\end{align}
Given that $\klam=1$ in the SM, it is instructive to expand \cref{eq:Zh_full} in terms of $(\klam-1)$, which is of $\order{1/\lamNP^2}$ in the SMEFT. This expansion can be done in several ways, in particular: 
\vspace{-3mm}
\begin{itemize}
    \item by including terms up to $\order{1/\lamNP^2}$:
    \begin{align}
    \frac{\sZh{\klam}}{\sZh{\klam=1}} &\simeq 1 + (\klam - 1) \left(C_1 + \frac{2\delta Z_h}{1 - \delta Z_h}\right)\;,
    \label{eq:Zh_nolambda4}
    \end{align}\vspace{-5mm}
    
    \item by including terms up to $\order{1/\lamNP^4}$:
    \begin{align}
        \frac{\sZh{\klam}}{\sZh{\klam=1}} \simeq 1
        + (\klam - 1) \left(C_1 + \frac{2\delta Z_h}{1 - \delta Z_h}\right)
        + (\klam - 1)^2 \left( \delta Z_h \frac{1 + 3\delta Z_h}{(1 - \delta Z_h)^2} + 2 C_1 \frac{\delta Z_h}{1 - \delta Z_h} \right)\;,
        \label{eq:Zh_with_C1_terms}
    \end{align}\vspace{-0.65cm}
    
    \item by including terms up to $\order{1/\lamNP^4}$, but only strictly one-loop contributions (\ie, not performing the resummation in the definition of $Z_{h,\klam}$):
    \begin{align}
        \frac{\sZh{\klam}}{\sZh{\klam=1}} \simeq 1 + (\klam - 1) \left( C_1 + 2\delta Z_h \right) + (\klam - 1)^2 \delta Z_h\;,
    \label{eq:Zh_stricly1L}
    \end{align}\vspace{-7mm}
    
    \item by including terms up to $\order{1/\lamNP^4}$, but excluding ``mixing'' terms, proportional to $C_1 \cdot {\delta Z_h}$:
    \begin{align}
        \frac{\sZh{\klam}}{\sZh{\klam=1}} \simeq 1 + (\klam - 1) \left( C_1 + 2\frac{\delta Z_h}{1 - \delta Z_h} \right)  + (\klam - 1)^2 \delta Z_h \frac{1 + 3\delta Z_h}{(1 - \delta Z_h)^2}\;,
        \label{eq:Zh_hepfit}
    \end{align}\vspace{-6mm}
    
    \item or by including terms up to $\order{1/\lamNP^6}$:
    \begin{align}
        \frac{\sZh{\klam}}{\sZh{\klam=1}} \simeq 1
        & + (\klam - 1) \left(C_1 + \frac{2\delta Z_h}{1 - \delta Z_h}\right)
        + (\klam - 1)^2 \left( \delta Z_h \frac{1 + 3\delta Z_h}{(1 - \delta Z_h)^2} + 2 C_1 \frac{\delta Z_h}{1 - \delta Z_h} \right) \nonumber \\
        & + (\klam - 1)^3 \left( 4\delta Z_h^2\dfrac{1 + \delta Z_h}{(1 - \delta Z_h)^3} + C_1 \delta Z_h\dfrac{1 + 3\delta Z_h}{(1 - \delta Z_h)^2} \right)\;.
        \label{eq:Zh_cubic}
    \end{align}\vspace{-6mm}
\end{itemize}

\cref{eq:Zh_hepfit} is used as the SMEFT prediction for $\sZh{}$ in the \hepfit package. The specific values for the $C_1$ coefficients at different \cm energies are taken from Ref.~\cite{DiVita:2017vrr}. All other single-Higgs observables at future circular \epem machines (namely, the $\epem\to\upnu\bar{\upnu}\Ph$ production cross-section and the Higgs partial widths) can also be parametrised as functions of \klam following \cref{eq:sigmaZh_smeft}, where only the $C_1$ terms are process-dependent. \hepfit includes these also following \cref{eq:Zh_hepfit}, with $C_1$ coefficients for the partial widths taken from Ref.~\cite{Degrassi:2016wml}. Apart from the one-loop contributions depending on \klam, \hepfit also implements the full leading-order SMEFT prediction for each~observable.

\begin{figure}[htpb]
    \centering
    \begin{subfigure}{0.32\textwidth}
        \centering
        \resizebox{!}{2.8cm}{% \resizebox{2.8cm}{!}{%
	\begin{tikzpicture}[scale=1.0]
		\setlength{\feynhandlinesize}{1.0pt}
		\setlength{\feynhanddotsize}{1.5mm}
		\begin{feynhand}
			\vertex [particle] (em) at (-2,+1.5) {\large ${e^{-}}$};
			\vertex [particle] (ep) at (-2,-1.5) {\large ${e^{+}}$};
			
			\vertex [particle] (Z) at (+2.5,+1.5) {\large ${Z}$};
			
			\vertex [particle] (h) at (+2.5,-1.5) {\large ${h}$};
			
			\vertex (veeZ) at  (-0.75,+0);
			\vertex (vZZh1) at (+0.5 ,+0);
			\vertex (vZZh2) at (+1.5 ,+0.5);
			\vertex (vhhh) at  (+1.5 ,-0.5);
			
			\propagator [fermion] (em) to (veeZ);
			\propagator [fermion] (veeZ) to (ep);
			
			\propagator [boson] (veeZ) to [edge label={\large ${Z}$}] (vZZh1);
			\propagator [scalar, blue!60!black] (vZZh1) to [edge label={\large ${A}$}] (vZZh2);
			
			\propagator [boson] (vZZh2) to (Z);
			
			\propagator [scalar, blue!60!black] (vZZh1) to [edge label'={\large ${H}$}] (vhhh);
			\propagator [scalar, blue!60!black] (vZZh2) to [edge label={\large ${H}$}] (vhhh);
			\propagator [scalar] (vhhh) to (h);
			
			\filldraw [blue!60!black] (vZZh1) circle (3pt);
			\filldraw [blue!60!black] (vZZh2) circle (3pt);
			\filldraw [blue!60!black] (vhhh) circle (3pt);
			
			%\vertex [red, below left=0.2mm of vhhh] {\large \kala};
			
		\end{feynhand}
	\end{tikzpicture}%
% }
}
                
        \vspace{-4mm}
        \caption{}
        \label{fig:ZH_IDM_diagrams:vertex1}
    \end{subfigure}\vspace{2mm}
    \begin{subfigure}{0.32\textwidth}
        \centering
        \resizebox{!}{2.8cm}{% \resizebox{2.8cm}{!}{%
	\begin{tikzpicture}[scale=1.0]
		\setlength{\feynhandlinesize}{1.0pt}
		\setlength{\feynhanddotsize}{1.5mm}
		\begin{feynhand}
			\vertex [particle] (em) at (-2,+1.5) {\large ${e^{-}}$};
			\vertex [particle] (ep) at (-2,-1.5) {\large ${e^{+}}$};
			
			\vertex [particle] (Z) at (+3,+1.5) {\large ${Z}$};
			
			\vertex [particle] (h) at (+3,-1.5) {\large ${h}$};
			
			\vertex (veeZ) at (-0.75,0);
			\vertex (vZZh) at (+1.0,0);
			\vertex (vhhh) at (+2.0,-0.75);
			
			\propagator [fermion] (em) to (veeZ);
			\propagator [fermion] (veeZ) to (ep);
			
			\propagator [boson] (veeZ) to [edge label={\large ${Z}$}] (vZZh);
			\propagator [boson] (vZZh) to (Z);
			\propagator [scalar, blue!60!black] (vZZh) to [quarter left, edge label={\large ${H}$}] (vhhh);
			\propagator [scalar, blue!60!black] (vZZh) to [quarter right, edge label'={\large ${H}$}] (vhhh);
			\propagator [scalar] (vhhh) to (h);
			
			\filldraw [blue!60!black] (vhhh) circle (3pt);
			\filldraw [blue!60!black] (vZZh) circle (3pt);
			
		\end{feynhand}
	\end{tikzpicture}%
% }
}
                
        \vspace{-4mm}
        \caption{}
        \label{fig:ZH_IDM_diagrams:vertex2}
    \end{subfigure}\vspace{2mm}
    \begin{subfigure}{0.32\textwidth}
        \centering
        \resizebox{!}{2.8cm}{\resizebox{2.5cm}{!}{%
\begin{tikzpicture}[scale=1.0]
	\setlength{\feynhandlinesize}{1.0pt}
	\setlength{\feynhanddotsize}{1.5mm}
	\begin{feynhand}
		\vertex [particle] (em) at (-2.25,+1.5) {\Large ${e^{-}}$};
		\vertex [particle] (ep) at (-2.25,-2) {\Large ${e^{+}}$};
		
		\vertex [particle] (Z) at (+3.5,+1.5) {\Large ${Z}$};
		
		\vertex [particle] (h) at (+3.5,-2.) {\Large ${h}$};
		
		\vertex (veeZ) at (-1.0,0);
		\vertex (vZZh) at (+1.0,0);
		\vertex (vhhh1) at (+1.8,-0.66);
		\vertex (vhhh2) at (+2.7,-1.33);
		
		\propagator [fermion] (em) to (veeZ);
		\propagator [fermion] (ep) to (veeZ);
		
		\propagator [boson] (veeZ) to [edge label={\Large ${Z}$}] (vZZh);
		\propagator [boson] (vZZh) to (Z);
		
		\propagator [scalar] (vZZh) to [edge label=\Large ${h}$] (vhhh1);
		
		\propagator [scalar, blue!60!black] (vhhh1) to [quarter left, looseness=1.5, edge label=\Large ${H}$] (vhhh2);
		\propagator [scalar, blue!60!black] (vhhh1) to [quarter right, looseness=1.5, edge label'=\Large ${H}$] (vhhh2);
		
		\propagator [scalar] (vhhh2) to (h);
		
		\filldraw [blue!60!black] (vhhh1) circle (4pt);
		\filldraw [blue!60!black] (vhhh2) circle (4pt);
		
	\end{feynhand}
\end{tikzpicture}%
}}
                
        \vspace{-4mm}
        \caption{}
        \label{fig:ZH_IDM_diagrams:WFR}
    \end{subfigure}
    \vspace{-7mm}
    \caption{Representative diagrams for the one-loop BSM corrections to $\epem\to\Zh$ in the IDM.}
    \label{fig:ZH_IDM_diagrams}
    \vspace{-0mm}
\end{figure}
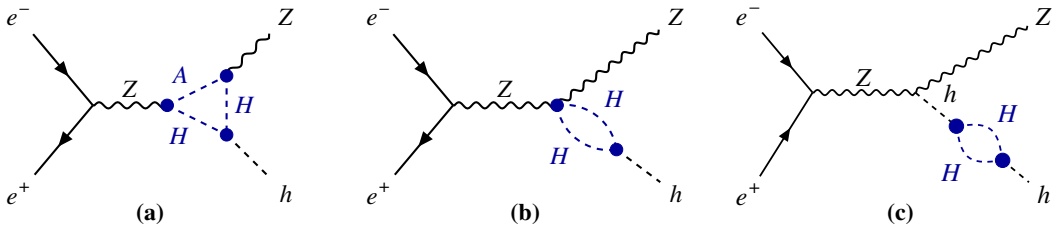

From the UV perspective, we perform a full one-loop IDM calculation for \lamhhh.
For the IDM predictions for the single-Higgs production observables, we include the SM-like diagrams with insertions of \klam, in accordance with the \hepfit implementation. For the $\epem\to\Zh$ process, we also evaluate the full one-loop momentum-dependent BSM vertex and external-leg corrections (BSM box corrections are suppressed by the electron Yukawa coupling and negligible); representative diagrams are depicted in \cref{fig:ZH_IDM_diagrams}. For the other single-Higgs observables, we incorporate the BSM corrections in terms of one-loop BSM predictions for the single-Higgs couplings $g_{\Ph XX}$.

\vspace{-0mm}
%%%%%%%%%%%%% figure %%%%%%%%%%%%%
%%%%%%%%%%%%% figure %%%%%%%%%%%%%

Instead of referring to the ratio $\sZh{}/\sZh{\text{SM}}$, in what follows it will be useful to introduce effective coupling modifiers to the \Zh cross-section, $\kZh{}$, defined by $\kZh{240} = \sqrt{\sZh{240}/(\sZh{240})^{\SM}}$ and $\kZh{365} = \sqrt{\sZh{365}/(\sZh{365})^{\SM}}$, where $\sZh{240}$ and $\sZh{365}$ refer to the $\epem\to\PZ\Ph$ production cross-sections at $\sqrt{s}=240\gev$ and $365\gev$, respectively.

\vspace{-0.5cm}
%%%%%%%%%%%%%%%%%%%%%%%%%%%%%%%%%%%%%%%%%%%%%%%%%%%%%%
%%%%%%%%%%%%%%%%%%%%%%%%%%%%%%%%%%%%%%%%%%%%%%%%%%%%%%
\section{Estimation of new nuisance parameters}
\label{sec:NPs}
\vspace{-0.3cm}

While EFT descriptions of New Physics benefit from a certain degree of model-independence, they also involve ambiguities in the choice of which operators to consider and where to truncate the $1/\lamNP$ expansion, given that it is not feasible to perform calculations up to all orders in $1/\lamNP$. In particular, the truncation of the EFT expansion introduces a source of theoretical uncertainty in the EFT predictions that is normally neglected in the evaluation of the \klam precision. In this section, we describe a procedure to estimate the size of these uncertainties for the case of $\sZh{}$ in SMEFT.

The $\sZh{}/\sZh{\text{SM}}$ ratio, or rather $(\kZh{})^2$, can be expressed in terms of \klam using \cref{eq:Zh_full}. This ratio can be written as an infinite series in terms of $(\klam-1)\propto 1/\lamNP^2$, which in turn can be truncated in different ways. In particular, \cref{eq:Zh_nolambda4,eq:Zh_with_C1_terms,eq:Zh_stricly1L,eq:Zh_hepfit,eq:Zh_cubic} lead to different expressions for \kZh{}, and the difference between these expressions serves as an indicator of the size of the theoretical uncertainties arising from the truncation of the SMEFT expansion in powers of $1/\lamNP$. Next-to-leading order (NLO) SMEFT contributions to $\sZh{}$ were only computed recently~\cite{Asteriadis:2024xuk,Asteriadis:2024xts}. Meanwhile, the public \hepfit implementation that we employ is restricted to leading-order contributions (of $\order{1/\lamNP^2}$) complemented only by the NLO effects directly dependent on \klam, as described in \cref{sec:sigmaZh}. The inclusion of NLO effects is not expected to have a drastic impact on the EFT truncation uncertainties.

Our procedure to estimate these theoretical uncertainties is as follows: firstly, we select a set of benchmark points (BPs) in the IDM parameter space, allowed by the current theoretical and experimental constraints. We then compute corresponding SMEFT Wilson coefficients, which we can then use to generate SMEFT predictions. More concretely, we note that, among the SMEFT operators relevant for $\klam$ at leading-order (\cref{eq:klam_smeft}), only $C_{\Phi}$ and $C_{\Phi\Box}$ are non-negligible in the IDM. Instead of performing a full matching from the IDM onto SMEFT, we can therefore follow a simpler procedure: first, we isolate the IDM contributions to \kZh{} corresponding to $C_{\Phi}$ and $C_{\Phi\Box}$; these are the SM-like diagrams with insertions of \klam (\cref{fig:ZH_klam_diagrams}) as well as the BSM external-leg corrections (\cref{fig:ZH_IDM_diagrams:WFR}). We also note that it is possible to define a bijective correspondence \hbox{$(C_\Phi, C_{\Phi\Box})\mapsto(\kZh{240},\kZh{365})$} by selecting any of the expressions in \cref{eq:Zh_nolambda4,eq:Zh_with_C1_terms,eq:Zh_stricly1L,eq:Zh_hepfit,eq:Zh_cubic} and adding the tree-level SMEFT contribution proportional to $C_{\Phi\Box}$, assuming all other Wilson coefficients to vanish. In particular, we consider \cref{eq:Zh_hepfit}, as implemented in \hepfit. This bijective correspondence can thus be inverted, such that, given IDM predictions for \kZh{240} and \kZh{365} (including only the aforementioned contributions), one directly obtains corresponding Wilson coefficients $C_{\Phi}$ and $C_{\Phi\Box}$ for the given IDM BP.

\begin{figure}[htpb]
    \centering
    \begin{subfigure}{0.49\textwidth}
        \centering
        \includegraphics[width=1.0\linewidth]{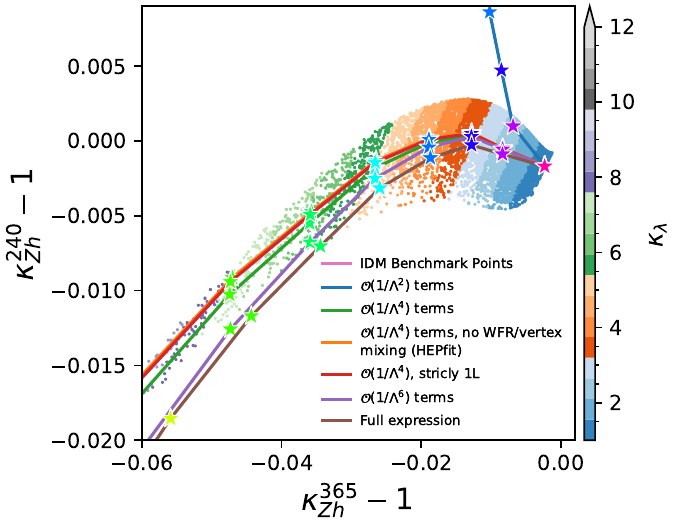}
        
        \vspace{-2mm}
        \caption{}
        \label{fig:firstNPs:curves}
    \end{subfigure}
    \begin{subfigure}{0.49\textwidth}
        \centering
         \includegraphics[width=0.86\linewidth]{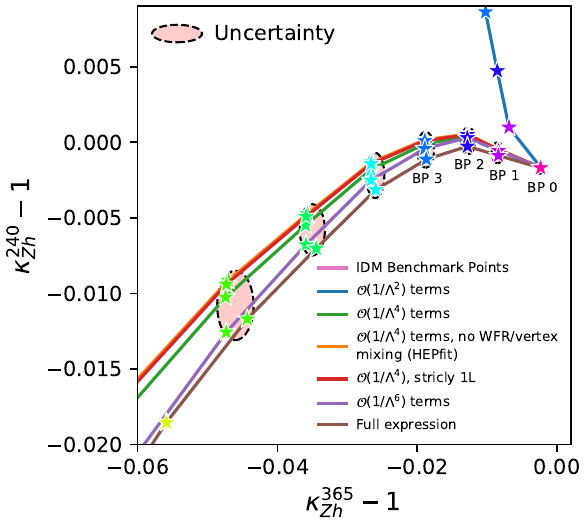}
         
         \vspace{-2mm}
        \caption{}
        \label{fig:firstNPs:ellipses_nobulk}
    \end{subfigure}
    \vspace{-4mm}
    \caption{(\subref{fig:firstNPs:curves}) The projection of the IDM parameter space onto $(\kZh{240},\kZh{365})$, with the IDM BPs and corresponding SMEFT predictions. The blue, green, red, orange, and purple curves correspond to \cref{eq:Zh_nolambda4,eq:Zh_with_C1_terms,eq:Zh_stricly1L,eq:Zh_hepfit,eq:Zh_cubic}, respectively, while ``Full expression'' (brown curve) refers to \cref{eq:Zh_full}. (\subref{fig:firstNPs:ellipses_nobulk}) The ellipses used for the initial estimates of the theoretical uncertainties.}
    \label{fig:firstNPs}
\end{figure}

\cref{fig:firstNPs:curves} shows the IDM predictions for \kZh{240} and \kZh{365} (excluding the BSM vertex corrections, in accordance with the previous paragraph) as scatter points, with their colours corresponding to the respective IDM prediction for \klam. Among these points, a set of benchmark scenarios was selected, each represented by stars filled with the same colour. Following the procedure outlined above, the values of $C_{\Phi}$ and $C_{\Phi\Box}$ were obtained for each IDM benchmark point, and these values were then used to generate different SMEFT predictions for \kZh{240} and \kZh{365} using \cref{eq:Zh_full,eq:Zh_nolambda4,eq:Zh_with_C1_terms,eq:Zh_stricly1L,eq:Zh_hepfit,eq:Zh_cubic}. The predictions obtained from each of these equations define a curve in \cref{fig:firstNPs:curves}, as indicated by the legend. Since \cref{eq:Zh_hepfit} is used to obtain the Wilson coefficients for each IDM BP, the corresponding curve (in orange) exactly matches the IDM curve (in pink), both of which are largely hidden by the red curve. The difference between the various SMEFT curves in \cref{fig:firstNPs:curves} provides an estimate of the theoretical uncertainties due to the truncation of the EFT expansion. This can be quantified by the ellipses around the set of SMEFT predictions for each given IDM BP shown in \cref{fig:firstNPs:ellipses_nobulk}, using their half-heights and half-widths as estimates for the uncertainties in \kZh{240} and \kZh{365}, respectively. For simplicity, we average both of these estimates over the $2<\klam<6$ range, to obtain a value independent of \klam. Note that the SMEFT predictions excluding all terms of order higher than $\order{1/\lamNP^2}$ (\cf\cref{eq:Zh_nolambda4}), shown in blue, deviate significantly from all other curves; therefore, we initially do not include this curve in the uncertainty estimation. Our initial estimates for the theoretical uncertainties in $(\kZh{240})^2$ and $(\kZh{365})^2$ are then $\NPI{240} = \NPI{365} = 0.107\%$. The inclusion of the strictly $\order{1/\lamNP^2}$ curve in the uncertainty evaluation leads instead to $\NPI{240}=1.15\%$ and $\NPI{365}=1.14\%$. These estimates are then used as the prior uncertainties for new nuisance parameters that we implement into \hepfit.

%%%%%%%%%%%%% figure %%%%%%%%%%%%%
%%%%%%%%%%%%% figure %%%%%%%%%%%%

\vspace{-0.6cm}
%%%%%%%%%%%%%%%%%%%%%%%%%%%%%%%%%%%%%%%%%%%%%%%%%%%%%%
%%%%%%%%%%%%%%%%%%%%%%%%%%%%%%%%%%%%%%%%%%%%%%%%%%%%%%
\section{Results and discussion}
\label{sec:results}
\vspace{-0.3cm}

\begin{figure}[htpb]
    \centering
    \begin{subfigure}{0.45\textwidth}
        \centering
         \includegraphics[width=1.0\linewidth]{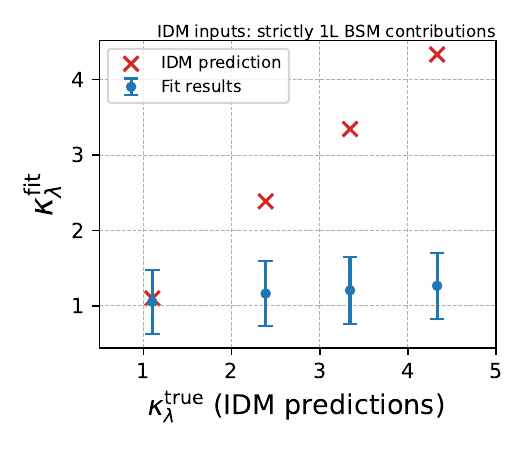}
         
         \vspace{-5mm}
        \caption{}
        \label{fig:results:pure1LBSM}
    \end{subfigure}
    \begin{subfigure}{0.46\textwidth}
        \centering
         \includegraphics[width=1.0\linewidth]{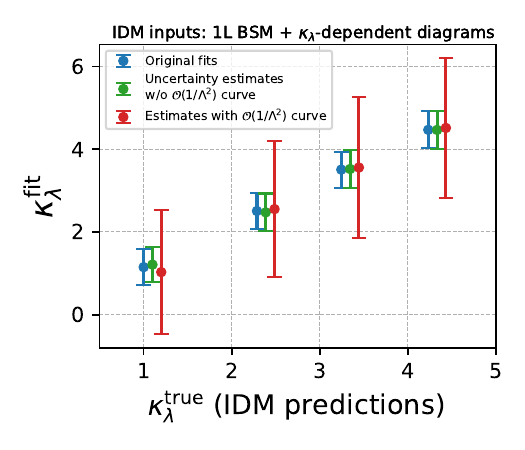}
         
         \vspace{-5mm}
        \caption{}
        \label{fig:results:NPs_comparison}
    \end{subfigure}
    \vspace{-3.5mm}
    \caption{(\subref{fig:results:pure1LBSM}) The results of the fits with strictly one-loop IDM inputs (blue), in comparison with the IDM predictions for \klam (red).
    (\subref{fig:results:NPs_comparison}) The global fit results for $\klam$ in the four IDM benchmark scenarios. The blue markers depict the original fit results without the new nuisance parameters, while the other colours illustrate the results using the two different sets of estimates for these parameters, with and without the inclusion of the strictly $\order{1/\lamNP^2}$ SMEFT curve (in red and green, respectively).}
    \label{fig:results}
\end{figure}

For the final selection of IDM BPs, all IDM contributions to \kZh{} are now included and all points are chosen to satisfy state-of-the-art experimental and theoretical constraints, including perturbative unitarity, boundedness-from-below of the Higgs potential, dark matter phenomenology, direct collider searches and EWPO constraints (see \eg Ref.~\cite{Braathen:2024ckk} for a recent review). We focus on four benchmark scenarios (illustrated in \cref{fig:firstNPs:ellipses_nobulk}), whose model parameters are shown in \cref{table:bp_parameters}. Following the procedure outlined in \cref{sec:method}, we perform global SMEFT fits using IDM ``Asimov'' pseudo-data. As a first step of our study, we generate fits using only strictly one-loop contributions in IDM inputs, \ie, we exclude the diagrams with insertions of \klam (\cref{fig:ZH_klam_diagrams}) from the fit inputs. \cref{fig:results:pure1LBSM} illustrates the resulting values of \klam extracted from the SMEFT fit, where a clear deviation can be observed from their respective IDM predictions. In fact, the global fit is seemingly unable to determine the presence of New Physics, as the extracted \klam values are consistent with the SM prediction in all benchmark scenarios. However, this study is arguably too naive, since the \klam-dependent contributions to \sZh{} correspond to a two-loop-level effect in the IDM. On the other hand, this first step of our study makes use of inputs at a consistent order in perturbation theory and clearly illustrates the significance of higher-order corrections to the determination of \klam.

%%%%%%%%%%%%%% table %%%%%%%%%%%%%%
\vspace{2mm}
{\renewcommand{\arraystretch}{1.2}
\begin{table}[htpb]
\footnotesize
\centering
\setlength{\tabcolsep}{3pt}
\begin{tabular}{|l|c|c|c|c|c|c|c|c|c|c|}
\hline
& $\mu_2^2\;[\si{\GeV^2}]$ & $\lambda_1$ & $\lambda_2$ & $\lambda_3$ & $\lambda_4$ & $\lambda_5$ & $m_H\;[\si{\GeV}]$ & $m_A\;[\si{\GeV}]$ & $m_{H^{\pm}}\;[\si{\GeV}]$ & $\kappa_\lambda$ \\ \hline\hline
BP 0 & \num[exponent-mode = scientific]{1.178e+06} & $0.2581$ & $6.213$ & $7.169$ & $-4.512$ & $-3.143$ & $1079$ & $1164$ & $1181$ & $1.1$ \\ \hline
BP 1 & \num[exponent-mode = scientific]{3.666e+05} & $0.2581$ & $3.084$ & $11.46$ & $-5.68$ & $-5.109$ & $622.2$ & $834.8$ & $845.1$ & $2.387$ \\ \hline
BP 2 & \num[exponent-mode = scientific]{3.922e+05} & $0.2581$ & $10.06$ & $14.19$ & $-6.974$ & $-6.407$ & $645.6$ & $897.3$ & $906.8$ & $3.345$ \\ \hline
BP 3 & \num[exponent-mode = scientific]{3.432e+05} & $0.2581$ & $8.985$ & $15.83$ & $-7.704$ & $-7.4$ & $604.3$ & $902.1$ & $907.2$ & $4.333$ \\ \hline
\end{tabular}
\vspace{-1.5mm}
\caption{The model parameters for the IDM benchmark points. We follow the conventions of Ref.~\protect\cite{Aiko:2023nqj}.}
\label{table:bp_parameters}
\end{table}}
\vspace{2mm}
%%%%%%%%%%%%%% table %%%%%%%%%%%%%%

\vspace{-2mm}
For more robust results, we again include the IDM vertex and external-leg corrections to \sZh{}, also including the diagrams with insertions of \klam (as described in \cref{sec:sigmaZh}). We perform this analysis with and without the inclusion of the new nuisance parameters defined in \cref{sec:NPs}. The results for \klam are illustrated in \cref{fig:results:NPs_comparison}. It is clear that using the initial estimates for \NPI{240} and \NPI{365} leads to little effect on the fit results, whereas the inclusion of the strictly $\order{1/\lamNP^2}$ curve in the estimates for the nuisance parameters leads to a significant decrease in the \klam precision. The latter result is not fully realistic, as the Higgs external-leg correction to \sZh{} (\cref{fig:ZH_klam_diagrams:WFR}), of $\order{1/\lamNP^4}$, is independent of the \cm energy and does not affect the central-value of \klam extracted from the global fit with \hepfit. However, this analysis highlights the sensitivity of the fit to generic effects of $\order{1/\lamNP^4}$, which are not fully taken into account in the current EFT analysis. This result therefore corresponds to a more consistent estimation of the theoretical uncertainties~from~the~SMEFT~truncation. \vspace{-1mm}

%%%%%%%%%%%%% figure %%%%%%%%%%%%%
%%%%%%%%%%%%% figure %%%%%%%%%%%%%

The results of this work illustrate the impact of theory uncertainties in the loop-level determination of \klam from single-Higgs observables at future \epem colliders. We mainly addressed here the uncertainties arising in the truncation of the EFT expansion. In the future, we plan to assess other potential sources of uncertainty, including those where the New Physics scenario assumed to be realised in nature departs from the decoupling limit in which SMEFT is most consistently applied, as well as uncertainties in the evaluation of the $C_1$ coefficients entering the SMEFT expressions for the single-Higgs observables. We stress that on the other hand the direct extraction of \klam from di-Higgs production observables at a future linear \epem machine will be much less susceptible to these sources of theory uncertainties, since \klam contributes to such process already at leading~order~\cite{Barklow:2017awn}.

\vspace{-0.3cm}
%%%%%%%%%%%%%%%%%%%%%%%%%%%%%%%%%%%%%%%%%%%%%%%%%%%%%%
%%%%%%%%%%%%%%%%%%%%%%%%%%%%%%%%%%%%%%%%%%%%%%%%%%%%%%
\acknowledgments
\vspace{-0.3cm} 
We thank Jorge de Blas for very helpful discussions. The work was made possible with the support of a scholarship from the German Academic Exchange Service (DAAD). J.B., J.L.\ and G.W.\ acknowledge support by the Deutsche Forschungsgemeinschaft (DFG, German Research Foundation) under Germany's Excellence Strategy --- EXC 2121 ``Quantum Universe'' --- 390833306. The work of P.B.\ and M.V.\ has been performed in the context of the Cluster EXC~3107 ``Color meets Flavor'' - 533766364 under Germany's Excellence Strategy. J.B.\ is supported by the DFG Emmy Noether Grant No.\ BR 6995/1-1. This work has been partially funded by the Deutsche Forschungsgemeinschaft 
(DFG, German Research Foundation) --- 491245950. 
The work of S.H.\ has received financial support from the grant PID2019-110058GB-C21 funded by
MCIN/AEI/10.13039/501100011033 and by ``ERDF A way of making Europe'', 
and in part by by the grant IFT Centro de Excelencia Severo Ochoa CEX2020-001007-S
funded by MCIN/AEI/10.13039/501100011033. 
S.H.\ also acknowledges support from Grant PID2022-142545NB-C21 funded by
MCIN/AEI/10.13039/501100011033/ FEDER, UE. 

\bibliographystyle{JHEP}
\bibliography{biblio.bib}

\end{document}